# Scalable CMOS-BEOL compatible AlScN/2D Channel FE-FETs

Kwan-Ho Kim[1], Seyong Oh[5], Merrilyn Mercy Adzo Fiagbenu[1], Jeffrey Zheng[2], Pariasadat Musavigharavi[1,2], Pawan Kumar[1], Nicholas Trainor[3], Areej Aljarb[4,7], Yi Wan[4], Hyong Min Kim[1], Keshava Katti[1], Zichen Tang[1], Vincent C. Tung[4,6], Joan Redwing[3], Eric A. Stach[2], Roy H. Olsson III[1]*, Deep Jariwala[1]*.

**Affiliations:**

[1]Department of Electrical and Systems Engineering, University of Pennsylvania, Philadelphia, PA, USA.

[2]Department of Materials Science and Engineering, University of Pennsylvania, Philadelphia, PA, USA.

[3]Department of Materials Science and Engineering, Pennsylvania State University, State College, PA, USA.

[4]Department of Physical Science and Engineering, King Abdullah University of Science and Technology, Thuwal, KSA.

[5]Querrey Simpson Institute for Bioelectronics, Northwestern University, Evanston, IL, USA.

[6]Department of Chemical System and Engineering, University of Tokyo, Tokyo, Japan

[7]Department of Physics, King Abdulaziz University (KAAU), Jeddah 23955-6900, Saudi Arabia

*Corresponding author. Email: dmj@seas.upenn.edu, rolsson@seas.upenn.edu.

**Abstract:** Intimate integration of memory devices with logic transistors is a frontier challenge in computer hardware. This integration is essential for augmenting computational power concurrently with enhanced energy efficiency in big-data applications such as artificial intelligence. Despite decades of efforts, reliable, compact, energy efficient and scalable memory devices are elusive. Ferroelectric Field Effect Transistors (FE-FETs) are a promising candidate but their scalability and performance in a back-end-of-line (BEOL) process remain unattained. Here, we present scalable BEOL compatible FE-FETs using two-dimensional (2D) $MoS_2$ channel and AlScN ferroelectric dielectric. We have fabricated a large array of FE-FETs with memory windows larger than 7.8 V, ON/OFF ratios of greater than $10^7$, and ON current density greater than 250 uA/um, all at ~80 nm channel lengths. Our devices show stable retention up to $2\times10^4$ secs and endurance up to $2\times10^4$ cycles in addition to 4-bit pulse programmable memory features thereby opening a path towards scalable 3D hetero-integration of 2D semiconductor memory with Si CMOS logic.

The generation of vast amounts of data by both ubiquitous electronic devices and the sensors within them, presents an acute need for computational power for high-speed and energy efficient data storage and processing. In conventional processor-centric computing, a processor core has to traverse various levels of memory at varying distances and speed, resulting in a data bottleneck and inefficient data handling[1,2,3]. Recently, compute-in-memory (CIM) architectures with vertically stacked, dense, high-efficiency and tightly integrated memory has been suggested to overcome such data handling bottlenecks[4,5]. In contrast with conventional CIM architectures, which are primarily front end of the line (FEOL) [i.e., memory is co-located with silicon logic transistors and peripheral circuits on the same layer], a memory array vertically stacked directly over the FEOL can provide unprecedented advantage in areal density and energy efficiency. It can also reduce latency[6]. At a single unit level this scheme requires a fast, reliable and low-energy non-volatile memory (NVM) device that can be easily integrated with the processing transistors without occupying precious space on the logic layer[7]. This drives the need for materials and devices compatible with back-end-of-line (BEOL) processing. Therefore, monolithic 3D (M3D) integration of NVM devices and Si complementary metal oxide semiconductor (CMOS) logic is desirable not only from the perspective of bringing memory closer to the processing unit, but this approach can further extend the Moore's Law via CIM architectures[8,9,10].

With recent advances in ferroelectric materials such as $Hf_xZr_{1-x}O_2$ (HZO), the FE-FET is considered as one of the most promising, compact and energy efficient NVM candidates for M3D integration, as it allows non-destructive read operation[6,11,12]. While HZO FE-FETs have made significant advances, ferroelectric properties superior to HZO have recently been discovered in aluminum scandium nitride (AlScN)[13,14]. AlScN not only enjoys a high remnant polarization ($P_r$) of > 110 μC/cm$^2$ – which is more than three times that of HZO – but also has a low deposition temperature of 350 °C (Fig. S1). This makes it a very attractive candidate for BEOL-compatible FE-FETs[15]. However, despite these attractive attributes, there has been no demonstration of scalable FE-FETs using an AlScN ferroelectric to date.

In this work, we demonstrate for the first time, a scalable array of BEOL-compatible 2D/3D heterogeneous FE-FETs having ~80 nm channel length using 45 nm thick AlScN with monolayer MoS$_2$ channels, all grown via a wafer-scalable processes. Three notable advances are highlighted here. First, by reducing the thickness and increasing the scandium (Sc) doping concentration of AlScN, we have confirmed that the memory window (MW) and switching voltage of the FE-FET can be largely controlled, bringing them closer to the operating voltage range of conventional FLASH memory devices[16,17]. Second, we have demonstrated the creation of an array of highly-scaled AlScN/MoS$_2$ FE-FETs with a channel length ($L_{CH}$) as small as 78 nm, and which have both a high on-current/off-current (ON/OFF) of $10^7$ and a current density of 252 μA/μm. Third, we show stable retention ($2\times10^4$ secs), endurance ($2\times10^4$ cycles) and pulse programmed 4-bit memory operation, which is critical to compete with conventional FLASH technology, in these highly scaled devices. Finally, we explore their potential for applications as artificial synapses with 7-bit states of conductance (G) in an artificial neural network simulation.

Fig. 1a shows the schematic of an MoS$_2$/AlScN FE-FET. We use 45 nm or 100 nm thick $Al_{1-x}Sc_xN$ ferroelectric dielectric films deposited on 4-inch Pt (111)/Ti/SiO$_2$/Si wafers. We highlight that the substrate temperature during the deposition of AlScN was maintained at 350 °C, a BEOL-compatible thermal budget. Large-area single-layer MoS$_2$ was used as the channel material of the FE-FETs. The large-area MoS$_2$ films were prepared via three different methods CVD 1[18], MOCVD[19], and CVD 2 (see methods) on 2-inch sapphire wafers and were transferred onto 45/100 nm $Al_xSc_{1-x}N$ films for device fabrication and testing (see methods for details). The device surface morphology and interface structure were confirmed through scanning electron microscopy (SEM) and cross-sectional transmission electron microscopy (X-TEM). As shown in Fig. 1b,c, the FE-FET has a channel width of 20 μm and $L_{CH}$ of 500 nm. This $L_{CH}$ is further aggressively scaled down to ~ 78 nm as shown below in Fig. 2a High-low magnification high-angle annular dark field STEM (HAADF-STEM) image of the semiconductor/dielectric interface, and phase contrast lattice image of MoS$_2$ and AlScN interface combined with elemental analysis (EDS mapping) shows a single-layer of MoS$_2$ on top of the crystalline AlScN. There is no evidence of an oxide layer on the AlScN (Fig. 1c-e). It should be noted that minimizing oxidation of the AlScN top surface which forms the interface with the semiconductor is important to avoid serious performance degradation of FE-FETs (See Fig. S2).

Next, the J-E hysteresis loops of as-sputtered Al$_{1-x}$Sc$_x$N samples were first measured on a metal/AlScN/metal structural capacitor to study its coercive field ($E_c$). A value of -4.5/5.1 MV cm$^{-1}$ were extracted under 10 kHz excitation (Fig. 1f). Polarization-dependent leakage was observed in the loop as in previous reports[15]. To minimize the effect of the leakage current and estimate accurate a $P_r$, positive-up, negative-down (PUND) pulsed measurements were carried out using short pulse width of 500 ns (Fig. S3). The PUND results showed a saturated $P_r^+$ and $P_r^-$ of around 135 μC/cm$^2$ obtained by ($P_r^+$ + $P_r^-$)/2. Detailed information about the J-E loop and PUND measurements can be found in the methods section.

All of the as-fabricated long-channel ($L_{CH}$ = 500 nm) FE-FETs (based on three different large-area MoS$_2$ and 100 nm Al$_{0.68}$Sc$_{0.32}$N film) show counterclockwise hysteric $I_D$-$V_G$ plots with a very large MW of about 18 V, a high ON/OFF of 10$^7$ and an ON-current density of 71 μA/μm ($W_{CH}$ = 20 μm) at $V_{DS}$ = 1V (Fig. 1g). Under the positive (negative) gate voltage above $E_c$, the FE polarization is switched in the direction pointing towards the channel (opposite of channel), and consequently electrons are accumulated (depleted) in the channel, causing a low threshold voltage (LVT) (high threshold voltage, HVT) state. For a low-energy consumption and M3D integration of the FE-FETs with Si-CMOS, the switching voltage must be reduced. One way to achieve this is by reducing the AlScN thickness and increasing Sc doping concentration[20]. As shown in Fig. 1h, the FE switching voltage for maximum MW is reduced from 20 V (for 100 nm thick Al$_{0.72}$Sc$_{0.28}$N) to 10 V (for 45 nm Al$_{0.68}$Sc$_{0.32}$N) and consequently the MW also reduces from 21 V to 7.8 V. The MWs are obtained by subtracting a (LVT) from a (HVT) that are extracted at a current level of ($W_m$/$L_m$)×10$^{-7}$ A, where $W_m$ and $L_m$ are the mask channel width and length, respectively. To corroborate the evidence of FE switching observed in the FE-FETs, we also compare the transfer curves between 50 nm SiO$_2$/MoS$_2$ FETs and 45 nm AlScN/MoS$_2$ FE-FETs that are fabricated using the same CVD MoS$_2$, fabrication process and same device dimension only except for the gate oxide. As shown in Fig. 1i, the transfer curve of SiO$_2$/MoS$_2$ FETs show clockwise hysteric loop that originates from charge trapping[21], while that of AlScN/MoS$_2$ FE-FETs have a counterclockwise hysteric loop. In addition, even if the sweep range of the gate voltage is 3 times narrower (-10 to 10 V), the current level and ON/OFF are approximately 10$^4$ and 5×10$^4$ times larger, respectively, in AlScN/MoS$_2$ FE-FET compared to SiO$_2$/MoS$_2$ FET. This observation confirms FE switching in the AlScN/MoS$_2$ structure. Fig. 1j shows output curves of the devices which demonstrate an ON-current density of 252 μA/μm at $V_{DS}$ of 3 V. To the best of our knowledge this is among the highest current density values obtained without any channel doping or contact resistance engineering in a 2D channel FET. We attribute this to the large polarization of the AlScN FE, further highlighting the utility of our device structure and materials selection.

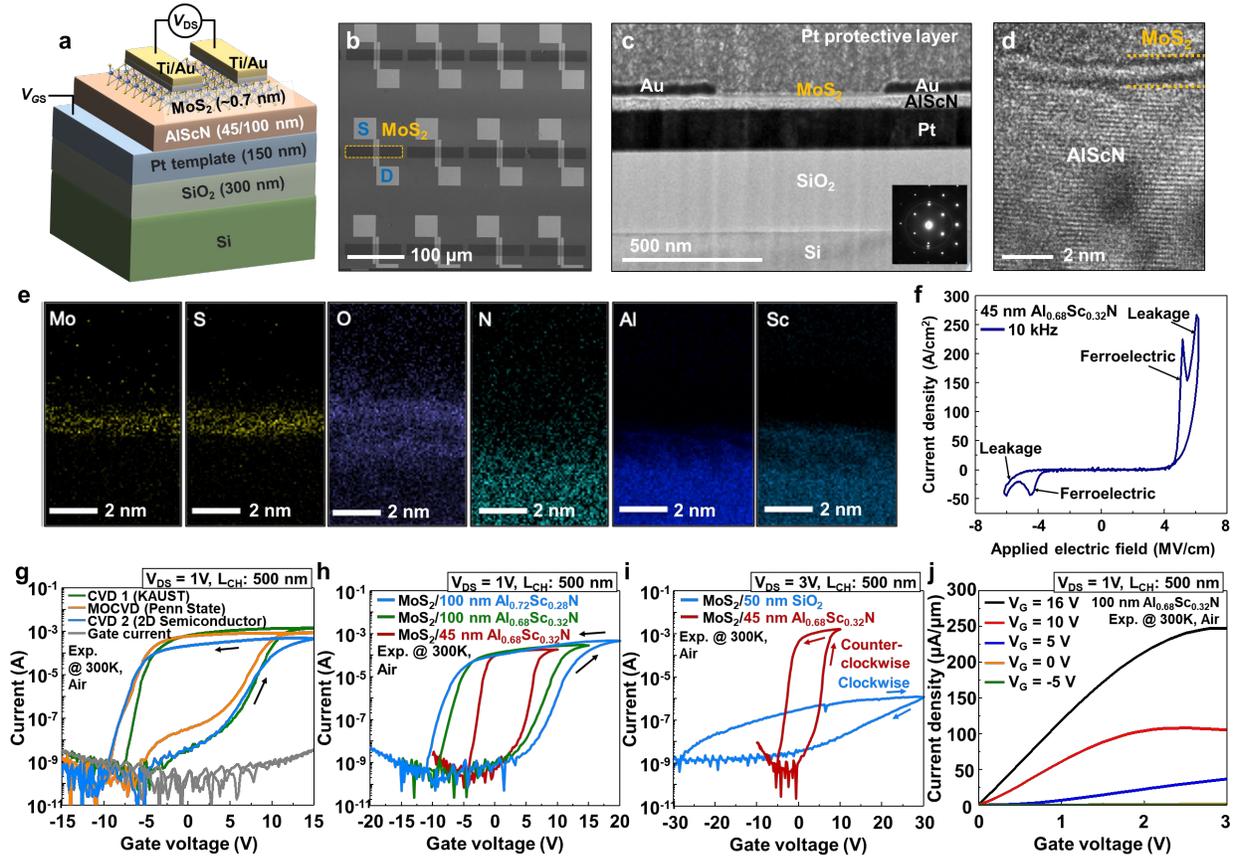

**Fig. 1. The AlScN/MoS$_2$ FE-FET device structure and electrical characteristics and the ferroelectric property of AlScN.** (a) Schematic of MoS$_2$/AlScN FE-FET. (b) SEM image of the FE-FET array (c) Cross-sectional STEM image of MoS$_2$/AlScN FE-FET. (d) Phase contrast lattice image of the MoS$_2$/AlScN interface. (e) Energy Dispersive X-Ray spectroscopic image of the FE-FET. (f) The J-E hysteresis loops of Al$_{0.68}$Sc$_{0.32}$N under 10 kHz. (g) Semilogarithmic scale transfer characteristics at room temperature of a representative AlScN/MoS$_2$ FE-FET based on the three different MoS$_2$ grown by CVD 1 (green), MOCVD (orange) and CVD 2 (blue), and (h) with changing the doping concentration of Sc (28% and 32%) and the thickness of AlScN (45nm and 100nm): 100 nm Al$_{0.72}$Sc$_{0.28}$N (blue curve), 100 nm Al$_{0.68}$Sc$_{0.32}$N (green curve) and 45 nm Al$_{0.68}$Sc$_{0.32}$N (red curve). (i) Semilogarithmic scale transfer characteristics at room temperature of a representative 45 nm AlScN/MoS$_2$ FE-FET (red) and 50 nm SiO$_2$/MoS$_2$ (blue), (j) Linear-scale output characteristics of a representative 100 nm Al$_{0.68}$Sc$_{0.32}$N/MoS$_2$ FE-FET at various gate voltage ($V_G$)

Next, we aggressively scale down the $L_{ch}$ of the FE-FET from 500 nm to 78 nm (Fig. 2a) while maintaining channel width. Further, we also extend our evaluation of device metrics over an array of devices. (See supporting information Fig. S4 and S5 for more details) Our devices show significant overlap in transfer characteristics and memory windows (Fig. 2b) making the case for viability at technology levels with further developments. It is worth noting that the FE-FETs maintain a large MW of 7.8 V and an ON/OFF > 10$^6$ even after the aggressive scaling of $L_{CH}$ and AlScN thickness. This is because the high $P_r$ of AlScN which noticeably keeps the OFF current low. For a fair assessment of our FE-FET performance, a comparison figure that compares normalized MW and on-state conductivity that are extracted from previous reports is included as shown in Fig. 2c. Since MW increases with the thickness of the ferroelectric, a normalized MW is the fair metric for comparison. Likewise, ON-state conductivity is also a normalized metric to the channel width/length and drain current. As can be seen from the figure, both the normalized MW and on-state conductivity of our device are among the highest compared to other 2D-channel FE-FETs. It is also noteworthy that both normalized MW and ON-state conductivity values are maintained even when the thickness of AlScN is reduced from 100 nm to 45 nm (the left and right red star in Fig. 2c corresponds to 45 nm and 100 nm AlScN, respectively). This suggests the possibility for further scaling down of AlScN thickness in future works without performance degradation. Statistical analysis of the MW (Fig. 2d) and ON/OFF ratios (Fig. 2e,f) are also shown. It is worth noting that the variation in HVT is larger than that in the LVT, and this is possibly related to the larger resistive leakage current of the P+ polarization[22]. Furthermore, the mean values of the ON/OFF$_{MAX}$ and ON/OFF at $V_G$ of 0 V (Fig. 2e,f) were observed to be $1.6 \times 10^6$ and $2.3 \times 10^5$, respectively. These variations in device-to-device transfer curves primarily originate from university laboratory/cleanroom-level device fabrication process and channel inhomogeneities which can be reduced with advanced foundry-based fabrication processes as well as with improved MoS$_2$ synthesis and AlScN deposition processes. The experimental I-V curves of our FE-FET devices are further verified via Technology Computer Aided Design (TCAD) simulations (Fig. S6). These simulations suggest a low level of fixed charge at the interface in our devices.

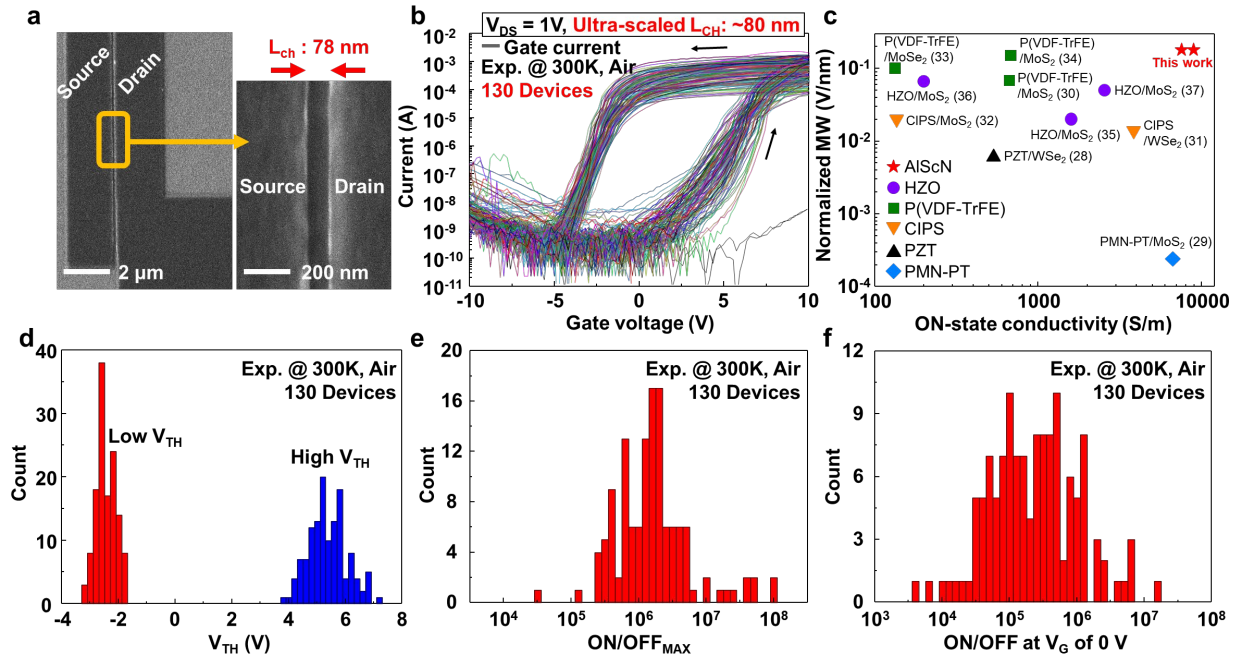

**Fig. 2. Array of ultra-scaled AlScN/MoS$_2$ FE-FETs.** (a) Magnified SEM image utilizing in-lens backscattered detector to confirm the channel length of the FE-FET. (b) Semilogarithmic scale transfer characteristics at room temperature of an array of 45 nm Al$_{0.68}$Sc$_{0.32}$N/MoS$_2$ FE-FETs with channel lengths of around 80 nm (total 130 devices). (c) Comparison of normalized memory window (MW) and ON-state conductivity from the reported 2D channel FE-FETs in the literature with various ferroelectrics. The left and right red star in the graph corresponds to 45 nm and 100 nm AlScN/MoS$_2$ FE-FETs, respectively (d) Distribution of low $V_{TH}$ (LVT) and high $V_{TH}$ (HVT),

and (e) ON/OFF current maximum (ON/OFF$_{MAX}$), and (f) ON/OFF current at V$_G$ of 0 V of the array of FE-FETs

Next, we study the voltage-pulse-induced FE switching of our FE-FETs by applying various pulse amplitudes and widths. Pulsed-voltage-induced FE switching is important since FE-FETs will operate in a circuit application based on pulsed programming and erasing of resistance states. As shown in Fig. 3a, after applying a programming (PRG) or erasing (ERS) pulse, we measure the transfer characteristics of our ultra-scaled FE-FET in narrow DC gate voltage sweeps from -5 to 5 V in which the voltage range is lower than the switching voltage (Fig. 3a). The V$_{TH}$ of our FE-FETs can be controllably changed by these pulses from the initial state (black) to the LVT (red) or HVT (blue). The width and amplitude of PRG and ERS pulses used in our study range from (500 ns, 34 V) to (40 ms, 12 V). It should be noted that, as in the FE HfO$_2$, there is a significant trade-off relation between a pulse width and a pulse amplitude for the FE switching i.e.[16]: a shorter pulse width requires a higher pulse amplitude and vice versa, and the relationship between the two is exponential (Fig. 3d). Not only programming and erasing, but also understanding and evaluating the non-volatile retention of these states is equally important. In our devices, the retention of both LVT/HVT and ON/OFF of the ultra-scaled FE-FET was measured at room temperature in air (Fig. 3b,c). Our FE-FETs exhibit stable retention characteristics showing large MW of > 3.6V and ON/OFF of around $2\times10^2$ (at V$_G$ of 0V) even after $2\times10^4$ secs. We confirmed that the stable retention is reproducible even with a shorter PRG and ERS pulse width (Fig. S10). These retention measurements further solidify the evidence of FE switching as there is no retention observed in SiO$_2$/MoS$_2$ FETs, even when the same CVD MoS$_2$ and fabrication process are used (Fig. S8). Aside from time dependent retention, our devices also exhibit stable switching endurance for ~20000 cycles while maintaining a large MW of > 5 V and ON/OFF of around $10^3$ when a pulse with 10 V amplitude and 40 ms width is used (Fig. 3e,f). It is noteworthy that both the MW and ON/OFF gradually increase up to ~ 5000 cycles. We attribute this as a signature of the wake-up effect in AlScN. The wake-up effect is widely known to occur in ferroelectrics in particular for FE HfO$_2$[23]. For AlScN the study of this effect is nascent. Therefore, further study of its polarization switching mechanism and dynamics is necessary and is a subject of our ongoing investigations. Beyond 5000 cycles, both MW and ON/OFF gradually decrease again and abruptly disappear in the end. Despite this, our endurance and retention studies are the longest and most comprehensive demonstrations in 2D material-based FETs and memory devices.

Pulsed-voltage-induced resistance and threshold voltage switching are important attributes of FE-FETs since these can be made tunable as a function of voltage amplitude and duration to induce partial switching of the FE domains underneath the channel. This property of partial switching is due to the stochastic nature of FE domain switching. We explore this attribute in our FE-FET memory devices in terms of multi-bit operation. To boost the effective data density per NVM cell, multi-bit operation is a key feature of modern memories. Although the device performance of individual FE-FETs has reached or surpassed FETs in FLASH technology[16], a multi-bit demonstration in FE-FETs is still in its infancy even for HfO$_X$ on Si CMOS devices and has never been demonstrated before for nitride ferroelectrics. The large MW of our FE-FETs is due to the large E$_C$ of AlScN: this is a favorable attribute for demonstration of multi-bit storage in BEOL compatible FE-FETs. Fig. 4a shows the successful demonstration of 2-bit operation measured from 30 ultra-scaled FE-FETs. It is worth nothing that the four memory states have a relatively tight distribution in the 30 measured devices, a key requirement for a scalable and reliable multi-bit memory technology. The different levels of V$_{THS}$ are obtained by applying a different number and amplitude of pulses (see supporting Fig. S11). As shown in Fig. 4b, the obtained 2-bit V$_{TH}$ states also show stable retention up to $10^3$ secs. By splitting the $V_{THS}$ more finely, even 4-bit operation can be demonstrated in our device as shown in Fig. 4c. Furthermore, we confirm that the 4-bit $V_{TH}$ states can also be programmed using shorter pulse widths of 1 μs at the expense of programming energy due to the higher pulse amplitude (Fig. S11). This multi-bit operation indicates that multiple FE domains are contained in the channel of the FE-FET, and the domains can be partially polarized by programming pulses[22,24,25]. To the best of our knowledge this is the first demonstration of multi-state programming in ferroelectric nitrides and in BEOL compatible FE-FETs at this scale. These results suggest the foundation for a scalable M3D integration of memory with logic.

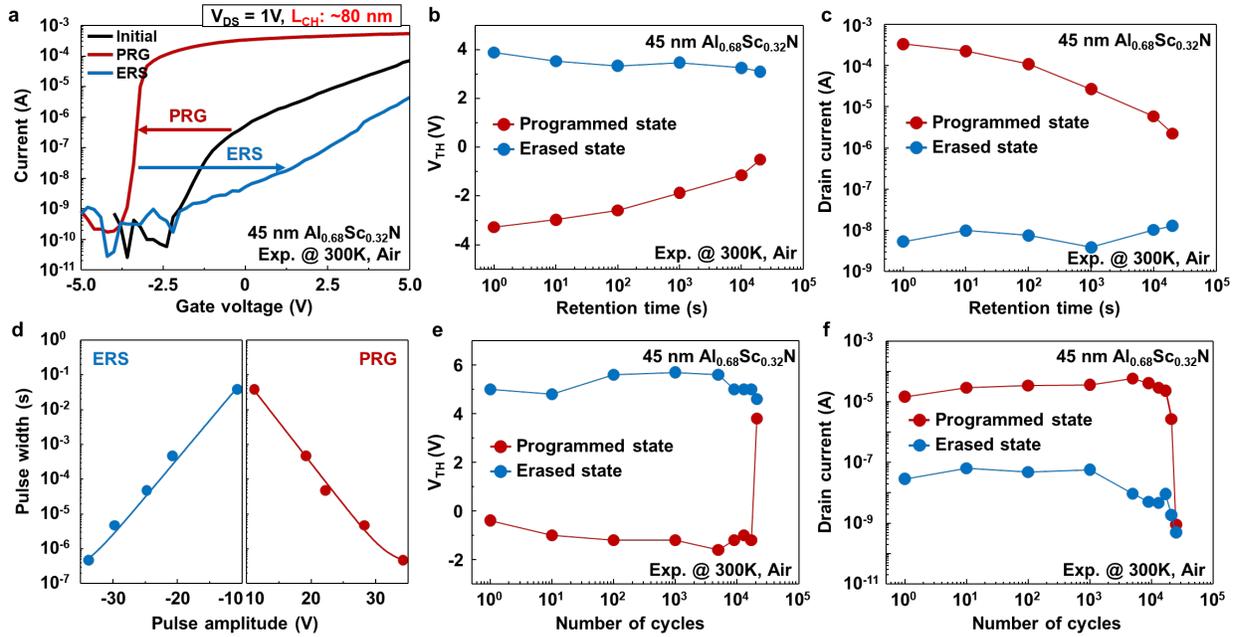

**Fig. 3. Electrical characterization of voltage pulse induced ferroelectric switching in the ultra-scaled FE-FET.** (a) Semilogarithmic scale transfer characteristics of the 45 nm AlScN/MoS$_2$ FE-FETs with channel length of around 80 nm after applying programming or erasing pulse with a width of 40 ms and an amplitude of ±12 V (b) Retention measurement of extracted threshold voltage ($V_{TH}$), and (c) Drain current at $V_G$ of 0 V, after applying programming/erasing pulse of 40 ms and ±12 V up to 20,000 secs. (d) Trade-off relation between pulse width and pulse amplitude for the ferroelectric switching. (e) Endurance measurements of the extracted threshold voltage ($V_{TH}$), and (f) Drain current at $V_G$ of 0 V, vs. the number of input pulse with 40 ms width and 10 V amplitude up to 20,000 times.

As a final demonstration, we show 7-bit (128) conductance states for pulse programmed operation of the FE-FET as an artificial synapse. Such fine programming of conductance suggests the presence of multiple FE domains in the channel, of which a small number switch stochastically with progressive numbers of pulses. Fig. 4d shows the synaptic weight update of our FE-FETs by each of 128 consecutive potentiation ($V_P$) and depression ($V_D$) pulses. Long term potentiation/depression (LTP/D) was observed when a pulse width of 150 μs and 15 V amplitude was applied to the gate at 4 kHz speed (see Fig. S12). This LTP/D behavior is observed to be reliable in extending the performance without degradation over at least 8000 programming pulses (Fig. 4e). In addition, a difference of excitatory postsynaptic output current (EPSC) (*27*) was observed to maintain up to at least 100 secs stably after applying various numbers of $V_P$ pulses (Fig. S13). Finally, we performed a multi-layer perceptron (MLP)-based artificial neural network (ANN) simulation using the open-source code "NeuroSimV3.0" (Fig. 4f)[26]. Here, the designed ANN consisted of 400 input, 100 hidden, and 10 output neurons, and each neuron was fully connected via artificial synapses that contained the device's nonlinear parameters. MNIST dataset of black and white handwritten digit patterns with a size of 20 × 20 were used for training (60,000) and testing (10,000). As a result, the maximum accuracy based on the LTP/D curve reached a very high accuracy of 94.26% (96.19% for software-based simulation).

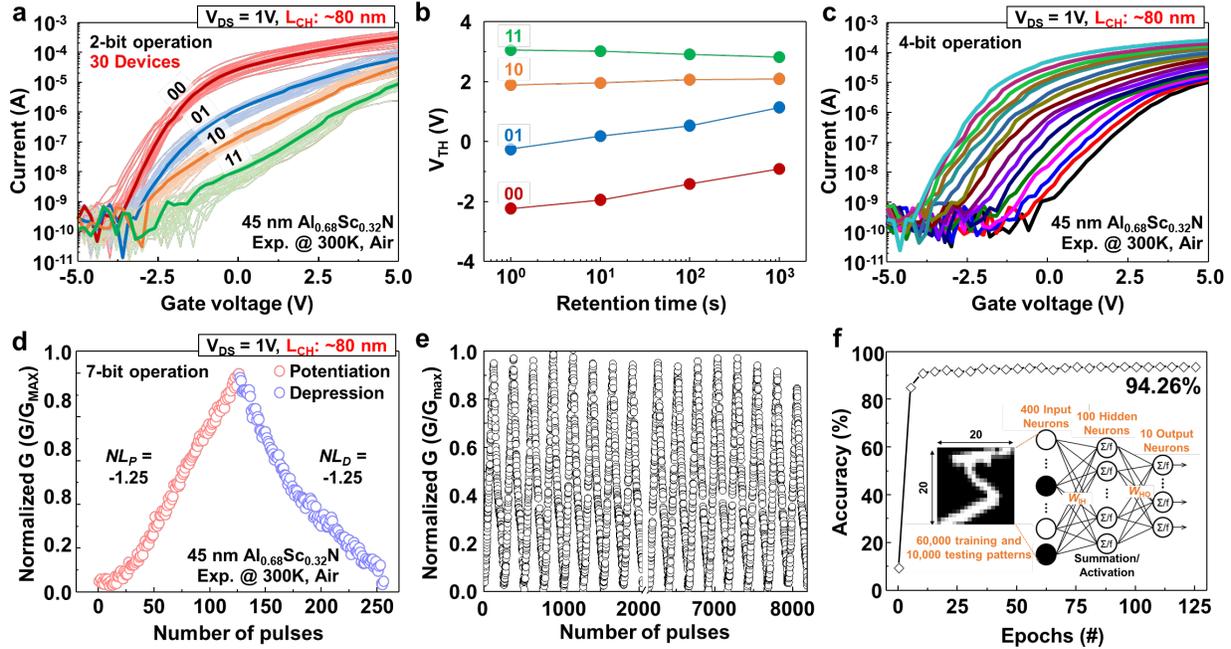

**Fig. 4. Multi-bit operation of the ultra-scaled FE-FET devices.** (a) 2-bit operation measured from 30 ultra-scaled FE-FETs after applying PRG or ERS pulses with an amplitude of 10 to 12 V and a width of 40 ms. (b) Retention of the 2-bit $V_{TH}$ states up to 1000 secs. (c) 4-bit operation measured from the ultra-scaled FE-FETs after applying PRG or ERS pulses with an amplitude of 9 to 12 V and a width of 40 ms. (d) Normalized 7-bit LTP/D curves obtained from the ultra-scaled FE-FET. Positive pulse with the 150 μs width and 15 V amplitude is used for LTP and negative pulse with the 150 μs and -1 to -15 V amplitude is used for LTD (e) Cycle-to-cycle variations of LTP/D curve for over 30 cycles (total over 8000 input pulses). (f) Recognition rate as a function of number of training epochs based on the LTP/D curve in Fig. 4d, and schematic illustration of multi-layer perceptron (MLP)-based artificial neural network (ANN) with size of 400 × 100 × 10 (Inset).

In summary, we have demonstrated NVM applications of a scalable and CMOS BEOL-compatible FE AlScN with < 1nm thick 2D channels down to ~80 nm channel length in array of devices. The stable memory performance of our devices combined with their scalability and low-temperature integration makes a promising case for vertical hetero-integration with Si CMOS logic transistors opening a new dimension to enhancing computational power and Moore's law with memory-centric and memory-enhanced computing in conjunction with conventional Si processors.

**References and Notes**

1. G. M. Marega et al., *Nature* **587**, 72-77 (2020).
2. Z. Wang et al., *Nat. Rev. Mater.* **5**, 173-195 (2020).
3. R. Yang et al., *Nat. Electron.* **2**, 108-114 (2019).
4. M. M. Shulaker et al., *Nature* **547**. 74-78 (2017).
5. A. Sebastian et al., *Nat. Nanotechnol.* **15**, 529-544 (2020).
6. S. Dutta et al., in *2020 IEEE International Electron Devices Meeting* (IEEE, 2020).
7. A. I. Khan et al., *Nat. Electron.* **3**, 588-597 (2020).
8. L. Tong et al., *Science* **373**, 1353-1358 (2021).
9. D. Akinwande et al., *Nature* **573**, 507-518 (2019).
10. D. K. Polyushkin et al., *Nat. Electron.* **3**, 486-491 (2020).
11. C. Sun et al., in *2021 IEEE Symposium on VLSI Technology* (IEEE, 2021).
12. K. Ni et al., *Nat. Electron.* **2**, 521-529 (2019).
13. D. Wang et al., *IEEE Electron Device Lett.* **41**, 1774-1777 (2020).
14. X. Liu et al., *Nano Lett.* **21**, 3753-3761 (2021).
15. D. Wang et al., *Phys. Status Solidi RRL* **15**, 2000575 (2021).
16. H. Mulaosmanovic et al., *Nanotechnol.* **32**, 502002 (2021).
17. T. Mikolajick et al., *J. Appl. Phys.* **129**, 100901 (2021).
18. A. Aljarb et al., *Nat. Mater.* **19**, 1300-1306 (2020).
19. A. Sebastian et al., *Nat. Commun.* **12**, 693 (2021).
20. S. Fichtner et al., *J. Appl. Phys.* **125**, 114103 (2019).
21. Y. Zhang et al., *Nat. Phys.* **5**, 722-726 (2009).
22. Y.-S. Liu, P. Su *IEEE Electron Device Lett.* **41**, 369-372.
23. C. Mart, *Appl. Phys. Lett.* **112**, 052905 (2018).
24. C. Li et al., in *2020 IEEE International Electron Devices Meeting* (IEEE, 2020), 20548637.
25. M. Lederer et al., *IEEE Trans. Electron Devices* **68**, 2295-2300 (2021).
26. X. Peng et al., in *2019 IEEE International Electron Devices Meeting* (IEEE, 2019), 19359366.
27. W. Xu et al., *Sci. Adv.* **2**, 1501326 (2016).
28. C. Ko et al., *Adv. Mater.* **28**, 2923-2930 (2016).
29. L. Xu et al., *ACS Appl. Mater. Interfaces* **12**, 44902-44911 (2020).
30. Y. T. Lee et al., *ACS Nano* **9**, 10394-10401 (2015).
31. X. Jiang et al., *ACS Appl. Electron. Mater.* **3**, 4711-4717 (2021).
32. M. Si et al., *ACS Nano* **12**, 6700-6705 (2018).

33. X. Wang et al., *2D Mater.* **4**, 025036 (2017).
34. L. Liu et al., *AIP Advances* **7**, 065121 (2017).
35. J. Xian et al., in *2021 Silicon Nanoelectronics Workshop* (SNW, 2021), 21046291.
36. K. Huang el al., *IEEE Electron Device Lett.* **10**, 1600-1603 (2020).
37. S. Zhang et al., *Nanoscale Research Lett.* **15**, 157 (2020).

**Acknowledgments:** This material is based upon work supported by the Defense Advanced Research Projects Agency (DARPA) TUFEN program under Agreement No. HR00112090046The work was carried out in part at the Singh Center for Nanotechnology at the University of Pennsylvania which is supported by the National Science Foundation (NSF) National Nanotechnology Coordinated Infrastructure Program (NSF grant NNCI-1542153). H.M.K., K.K. and D.J. acknowledge support from Penn Center for Undergraduate Research and Fellowships. The authors gratefully acknowledge use of facilities and instrumentation supported by NSF through the University of Pennsylvania Materials Research Science and Engineering Center (MRSEC) (DMR-1720530). P.K., E. A. S. and D. J. also acknowledge partial support from NSF DMR Electronic Photonic and Magnetic Materials (EPM) core program (Grant No. DMR-1905853)as well as the University of Pennsylvania Laboratory for Research on the Structure of Matter, a Materials Research Science and Engineering Center (MRSEC) supported by the National Science Foundation (No. DMR-1720530). A.A., Y.W., and V.T. are indebted to the support from the King Abdullah University of Science and Technology (KAUST) Solar Center and Office of Sponsored Research (OSR) under Award No: OSR-2018-CARF/CCF-3079. The MOCVD grown $MoS_2$ monolayer samples were provided by the 2D Crystal Consortium-Materials Innovation Platform (2DCC-MIP) facility at the Pennsylvania State University, which is funded by the NSF under cooperative agreement no. DMR-1539916.

**Author contributions:** D.J. and K.-H.K. conceived the idea. K.-H.K. designed experiments and performed device fabrication and characterization of the samples. K.-H.K. and D.J. wrote the manuscript. M.M.A.F. and K.-H.K. conducted P-E loop and PUND measurements and R.H.O. supervised them. J.Z. and K.-H.K. performed sputtering to prepare various AlScN substrates and R.H.O. supervised them. P.M. and P.K. performed transmission electron microscopy and scanning electron microscopy characterization respectively. P.K. made cross-sectional lamella for subsequent TEM observation. E.A.S. supervised the microscopy efforts. N.T. prepared 2-inch wafer scale $MoS_2$ using MOCVD and J.R. supervised it. CVD-based 2-inch wafer scale $MoS_2$ was prepared by A.A. and Y.W. and V.T. supervised them. K.-H.K. and P.K. performed wet transfer of $MoS_2$ on AlScN and $SiO_2$. S.O. and K.K. contributed to the MLP-based artificial neural network simulation and TCAD simulation. K.-H.K. and H.M.K. performed electrical measurements of 130 FE-FET array. Z.T. shared his recipe for the electron-beam lithography recipe. All authors contributed to the discussion and analysis of the results.

**Competing interests:** A sub-set of the authors (D.J., R.H.O., K.-H.K, T.O., E.A.S. and J.Z.) have filed a patent based on this work. The authors declare no other competing interests.

**Data and materials availability:** All data needed to evaluate the conclusions of this study are present in the paper or the supplementary materials.

# Supplementary Materials for

## Scalable CMOS-BEOL compatible AlScN/2D Channel FE-FETs

Kwan-Ho Kim, Seyong Oh, Merrilyn Mercy Adzo Fiagbenu, Jeffrey Zheng, Pariasadat Musavigharavi, Pawan Kumar, Nicholas Trainor, Areej Aljarb, Yi Wan, Hyong Min Kim, Keshava Katti, Zichen Tang, Vincent C. Tung, Joan Redwing, Eric A. Stach, Roy H. Olsson III*, Deep Jariwala*.

Correspondence to: dmj@seas.upenn.edu, rolsson@seas.upenn.edu

**Affiliations:**

[1]Department of Electrical and Systems Engineering, University of Pennsylvania, Philadelphia, PA, USA.

[2]Department of Materials Science and Engineering, University of Pennsylvania, Philadelphia, PA, USA.

[3]Department of Materials Science and Engineering, Pennsylvania State University, State College, PA, USA.

[4]Department of Physical Science and Engineering, King Abdullah University of Science and Technology, Thuwal, KSA.

[5]Querrey Simpson Institute for Bioelectronics, Northwestern University, Evanston, IL, USA.

[6]Department of Chemical System and Engineering, University of Tokyo, Tokyo, Japan

[7]Department of Physics, King Abdulaziz University (KAAU), Jeddah 23955-6900, Saudi Arabia

*Corresponding author. Email: dmj@seas.upenn.edu, rolsson@seas.upenn.edu.

## Materials and Methods

Synthesis of MoS₂ by CVD 1.

The monolayer MoS$_2$ films were synthesized on sapphire substrates in a 2-inch furnace using a conventional chemical vapor deposition (CVD) approach, which was adopted from the previous work[1]. In essence, high-purity MoO$_3$ powders (Sigma-Aldrich, 99.9%) and S powders (Sigma-Aldrich, 99.99%) were used as the precursor. The MoO$_3$ powders were placed in a ceramic crucible located at the center of the furnace, while the S powders were annealed by a heating tape at the upstream side. During the reaction, the furnace was heated to 800°C and the heating tape was heated to 140°C, in the meantime, 70 sccm of Ar gas was induced into the furnace to transport the precursors to the downstream side where the sapphire substrates were located. The reaction was maintained for 15 min at 30 torr.

Synthesis of MoS₂ by MOCVD.

Fully coalesced MoS$_2$ monolayers were also grown on 2-inch c-plane sapphire substrates using a horizontal MOCVD system from CVD Equipment. Mo(CO)$_6$ and H$_2$S were served as the Mo and S sources, respectively, with the former supplied from a stainless steel bubbler maintained at 10°C and 650 torr. During the growth, $1.1\times 10^{-3}$ sccm of Mo(CO)$_6$, 400 sccm H$_2$S and 4100 sccm of H$_2$ were feed into the reactor for 12 minutes, which was held at 900 °C and 50 torr. After the growth, the film was annealed for 10 minutes under H$_2$S before cooling down.

Synthesis of MoS₂ by CVD 2.

For sulfurization H$_2$S (g), H$_2$ (g), and MoO$_{3-x}$ (g) was used at around 690 °C. The temperature has various steps to initiate nucleation followed by enhancing lateral growth. Sapphire samples are subject to even higher temperature treatments before the growth to ensure predictable nucleation characteristics.

AlScN deposition.

The depositions of 45 and 100 nm AlScN were performed on 4-inch Pt (111)/Ti/SiO$_2$/Si wafers via 150 kHz pulsed DC co-sputtering with 20 sccm N$_2$ flow under $8.3 \times 10^{-4}$ mbar in an Evatec CLUSTERLINE® 200 II pulsed DC sputtering system. The chamber temperature was maintained at 350 °C, a BEOL compatible thermal budget.

PUND and P-E loop measurement.

For the positive-up negative-down (PUND) and P-E loop measurement, MFM capacitors (Pt/AlScN/Al) were used. The PUND measurements were taken on 20 um-radii capacitors using the Keithley 4200A-SCS analyzer. The voltage waveform of the PUND test consists of four monopolar pulses of pulse width 500 ns, rise or fall times of 140 ns, and delay (interval between subsequent pulses) of 1 us. Voltages (positive or negative) were applied to the bottom electrode of the MFM capacitors and current was sensed from the top electrode to minimize current transients in the data due to parasitic capacitances. The Keithley analyzer integrates the current measured in each pulse and returns the associated change in charge, which is related to polarization through a division by the area of the capacitor. In post-measurement processing using MATLAB, the remnant polarization for applied positive voltages (that is the P and U pulses) was obtained from the PUND measurements by subtracting the U polarization from the P polarization, while the remnant polarization for applied negative voltages (that is the N and D pulses) was obtained by subtracting the D polarization from the N polarization. The motivation behind such a measurement is first that switching from an upward (saturated) polarization state to a downward (saturated) polarization state requires a change in polarization of $-2\times P_r$, where $P_r$ is the average remnant polarization of the ferroelectric material, and the reverse requires a change in polarization of $+2\times P_r$. The second motivation is that by subtracting the U pulse from the P pulse, or the D pulse from the N pulse, one can reduce the contribution of leakage current—which becomes significant as applied voltages approach the breakdown limit—as well as of capacitive current from the measured polarization of the ferroelectric material.

The P-E loop measurements were performed on 20 um-radii capacitors using a Radiant Precision Premier II ferroelectric tester. The applied voltage utilized the standard bipolar profile: the voltage was ramped monotonically (linearly) from 0V to the maximum voltage, $V_{MAX}$, and then decreased monotonically to $-V_{MAX}$ before being increased back to 0V. The period of the voltage profile was 0.1ms, corresponding to a measurement frequency of 10kHz. The current density-vs-voltage characteristic was then extracted from the polarization-vs-voltage data by dividing the change in polarization at successive measurement points by the time interval between those measurements. The electric field was then obtained by dividing the applied voltage by the thickness (45 nm) of the MFM capacitors.

Fabrication of the large area $MoS_2$/AlScN FE-FET.

First, large-area $MoS_2$ having a size of around 1 $cm^2$ grown on sapphire was transferred by a wet transfer method. The $MoS_2$ transferred sample was dried in the glove box for a day. Then, the sample was coated using "PMMA A4 and PMMA A8" or "PMGI SF 5S and ZEP 520A" followed by source/drain (S/D) patterning by electron beam lithography (EBL). After develop using MIBK or PMGI 101A, 10 nm Ti and 30 nm Au were deposited as the S/D contact metal and pad metal, respectively, using electron beam evaporation. Samples were immersed in acetone or remover 1165 for approximately 20 mins, gently shaken to lift the metal, then rinsed with IPA and D.I. water. Next, in order to define channel area, a second patterning by photolithography was done after coating photoresists (LOR 3A and S1813), followed by develop with AZ-MIF-300. Finally, the exposed area of $MoS_2$ was etched using oxygen reactive ion etching (RIE).

Electrical measurement of the FE-FET.

Electrical measurements were performed in air at ambient temperature in a Lakeshore probe station using a Keithley 4200A semiconductor characterization system. For the DC I-V characteristics the SMU connection was used, while both PMU and SMU connections were used for the voltage-pulse-induced ferroelectric switching measurements.

S/TEM and SEM Characterization.

Scanning/transmission electron microscopy (S/TEM) characterization and image acquisition were carried out on a JEOL F200 operated at 200 kV accelerating voltage. Energy Dispersive X-ray Spectroscopy (EDS) analysis was performed using a JEOL NEOARM TEM operated at a voltage of 200 kV, with a point resolution better than 0.08 nm. This is a powerful technique, which can detect differences in composition on the atomic scale. All of the captured STEM images were collected/calculated using Digital Micrograph software (DM, Gatan Inc., USA). BrightBeam Scanning electron microscopy (SEM) was performed inside the dual-beam plasma FIB system (TESCAN S8000X). SEM operated at 5keV utilized in-lens backscattered MD detector to capture the high-resolution image for better analysis of ~80 nm channel length devices.

TEM/STEM Sample Preparation.

The TEM cross-sectional sample was prepared by a $Xe^+$ plasma-focused Ion Beam (TESCAN S8000X PFIB-SEM) system. The sample surface was coated with electron beam and ion beam deposited Pt protection layers to prevent damaging top surfaces and heating effects during FIB milling. FIB lamella was milled at 30keV and further in-situ lift-out technique was used with help of Kleindiek probe manipulator. Final thinning and cleaning of lamella was performed at 10keV and 5keV respectively.

Wet transfer.

$MoS_2$ grown on 2-inch wafer scale was first cut to 1 $cm^2$ size. The 1 $cm^2$ size $MoS_2$ was coated by PMMA 4 with 2000 RPM and 60 secs. After soaking the sample in 90° water for about 10 mins, potassium hydroxide (KOH) was used to separate $MoS_2$ from the sapphire. The detached $MoS_2$ was floated on D.I. water for about 15 mins to clean the (KOH). Finally, the $MoS_2$ was transferred on the AlScN in the D.I. water.

$MoS_2$ etching using RIE.

For the channel area define, $MoS_2$ was etched using March Jupiter II. The power of plasma was 100 W, the $O_2$ gas flow was 450 sccm and the run time was 30 secs.

| Ferroelectric materials | HZO | AlScN |
|---|---|---|
| Remanent polarization (µC/cm$^2$) | 5 - 35 | 75 - 135 |
| Coercive field (MV/cm) | 1 - 2 | 3 – 6.5 |
| Dielectric constant | 27 - 33 | 10 - 17 |
| Processing temperature (°C) | ≥ 400 | 350 |

**Fig. S1. Comparison table of the ferroelectric properties of AlScN and HZO**.

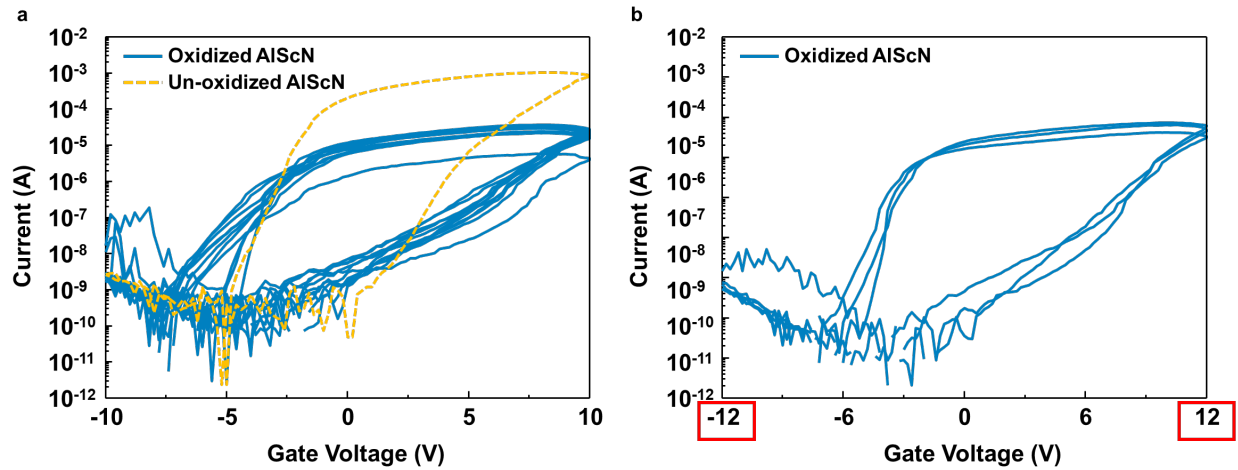

**Fig. S2. Transfer curves comparison between oxidized AlScN and un-oxidized AlScN.** (a) Transfer curve obtained from -10 to 10 V gate voltage sweep, and (b) -12 to 12 V. When oxidized AlScN was used as a ferroelectric gate oxide, both ON/OFF and on current level was significantly degraded as well as switching voltage for the maximum memory window increased 2 V because of the larger voltage drop across the oxide layer.

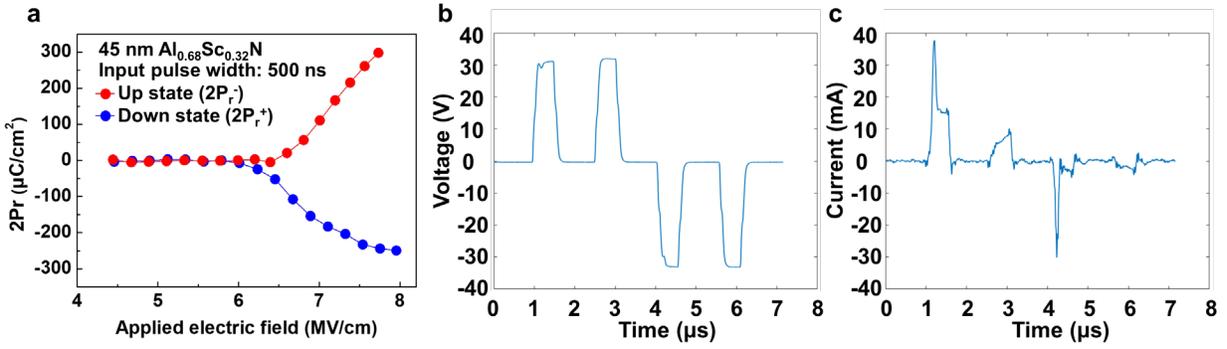

**Fig. S3. PUND measurements** (a) The $P_r$ of $Al_{0.68}Sc_{0.32}N$ calculated from the PUND measurements applying a pulse of 500 ns width and various voltage amplitude. (b) PUND voltage and (c) current densities profile showing ferroelectric switching within around 200 ns of the onset of the voltage switching pulse with a 140 ns rise and fall time.

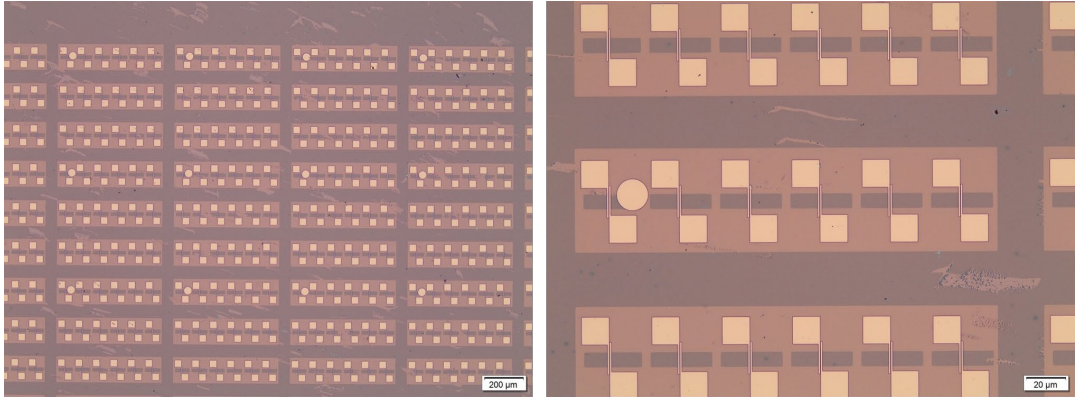

**Fig. S4. Optical images of the array of FE-FETs.**

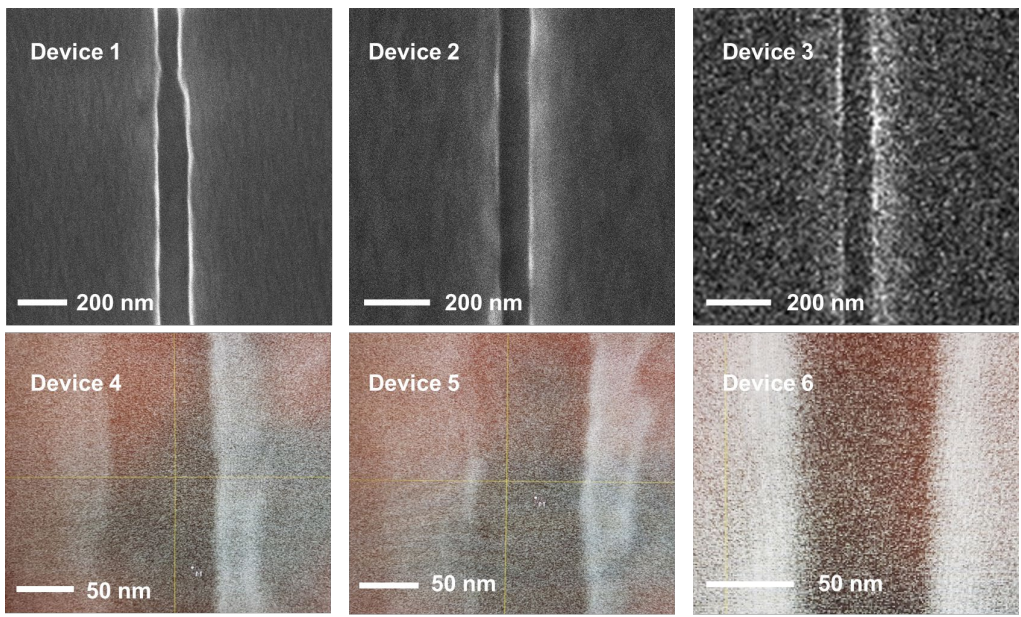

**Fig. S5. SEM images of 6 ultra-scaled FE-FETs.** All devices show the channel length of approximately 80 nm.

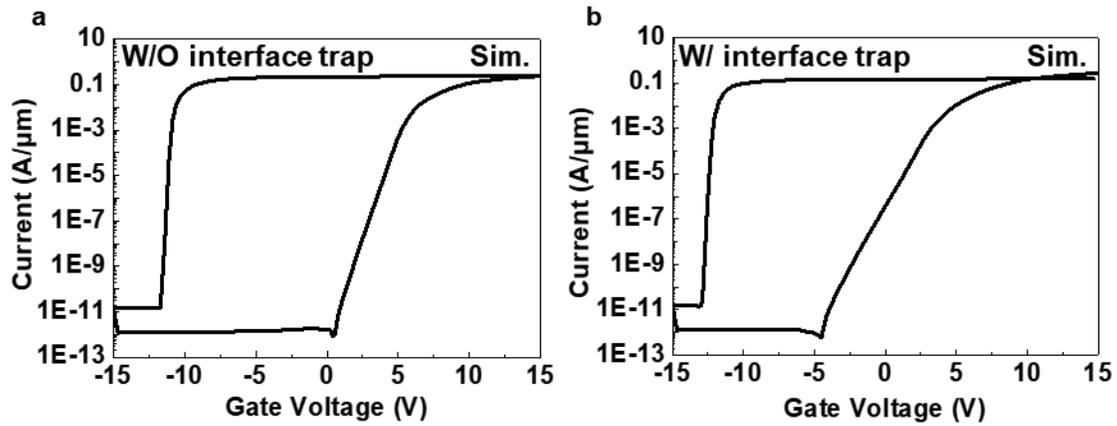

**Fig. S6. TCAD simulations of FE-FETs with AlScN of 45 nm and single-layered MoS₂,** (a) without interface trap and (b) with interface trap, showing counterclockwise transfer curve hysteresis loops. Channel length and width of the device were set to 28 nm and 20 μm, respectively, in the simulation. For the FE-FET without fixed charge density ($Q_F$) and with $Q_F$ of $10^{14}$/cm², and it showed strong agreement with experiments when a $Q_F$ of $10^{14}$/cm² at the MoS₂/AlScN interface is not considered, indicating a low level of $Q_F$ in our devices.

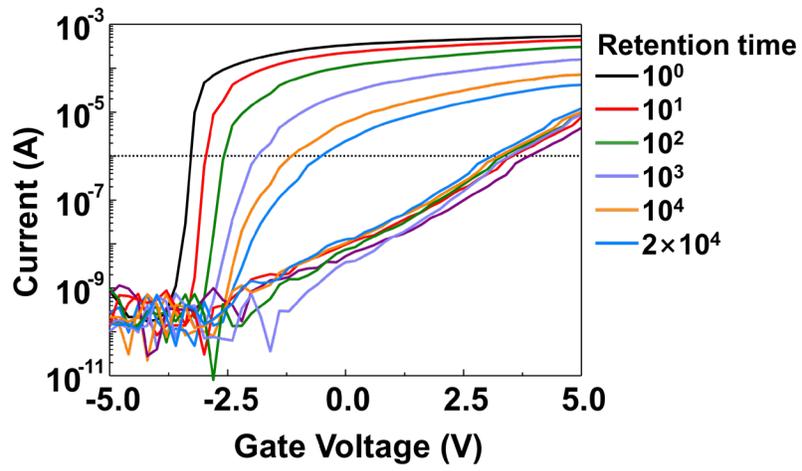

**Fig. S7.** Semilogarithmic scale transfer characteristics of the 45 nm AlScN/MoS$_2$ FE-FETs with L$_{CH}$ of around 45 nm according to the retention times. The amplitude and width of PGR pulse is 12 V and 40 ms, respectively.

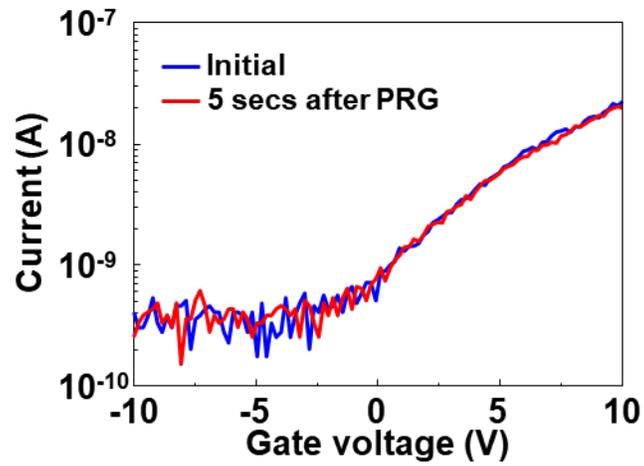

**Fig. S8. Semilogarithmic scale transfer characteristics of the 50 nm SiO$_2$/MoS$_2$ with L$_{CH}$ of around 500 nm before and 5 secs after PRG.** Before and after applying PRG pulse with 30 V and 40 ms, a narrow range of DC gate voltage (-10 to 10 V) is swept. However, a retention is not observed in the SiO$_2$/MoS$_2$ FET.

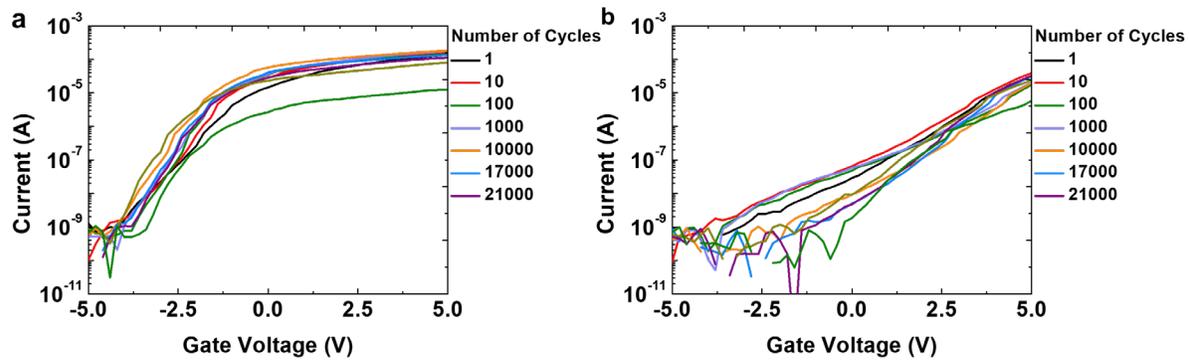

**Fig. S9.** Semilogarithmic scale transfer characteristics of the 45 nm AlScN/MoS$_2$ Fe-FETs with L$_{CH}$ of around 45 nm according to the number or cycles.

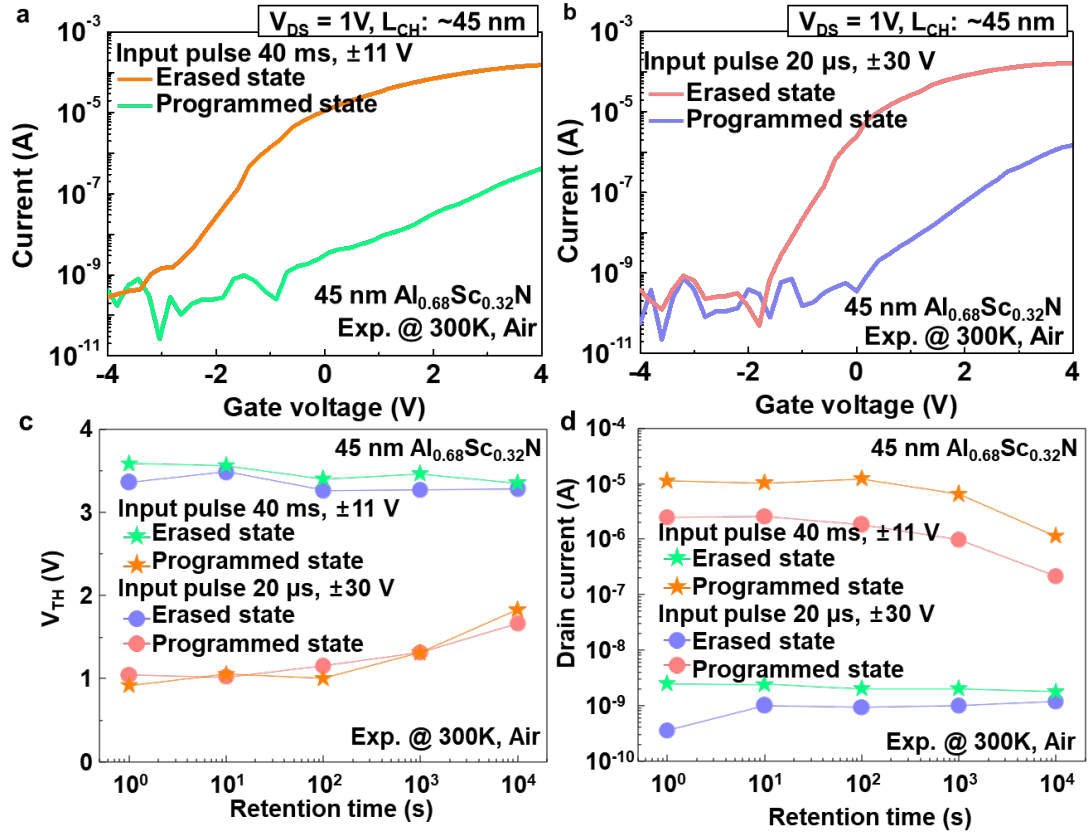

**Fig. S10. Semilogarithmic scale transfer characteristics of the 45 nm AlScN/MoS$_2$ Fe-FETs with L$_{CH}$ of around 45 nm after applying programming or erasing pulse with** (a) a width of 40 ms and an amplitude of +11 V or -11 V, (b) a width of 20 μs and an amplitude of +30 V or -30 V, and (c) and (d) their retention up to 10$^4$ secs. Here, the memory window is calculated using different V$_{TH}$ extraction method (linear extrapolation method) from the Fig. 3B of the manuscript. This was because the range of gate voltage sweep is narrower (-4 to 4 V), and thus the current level does not reach to $(W_m/L_m) \times 10^{-7}$ A.

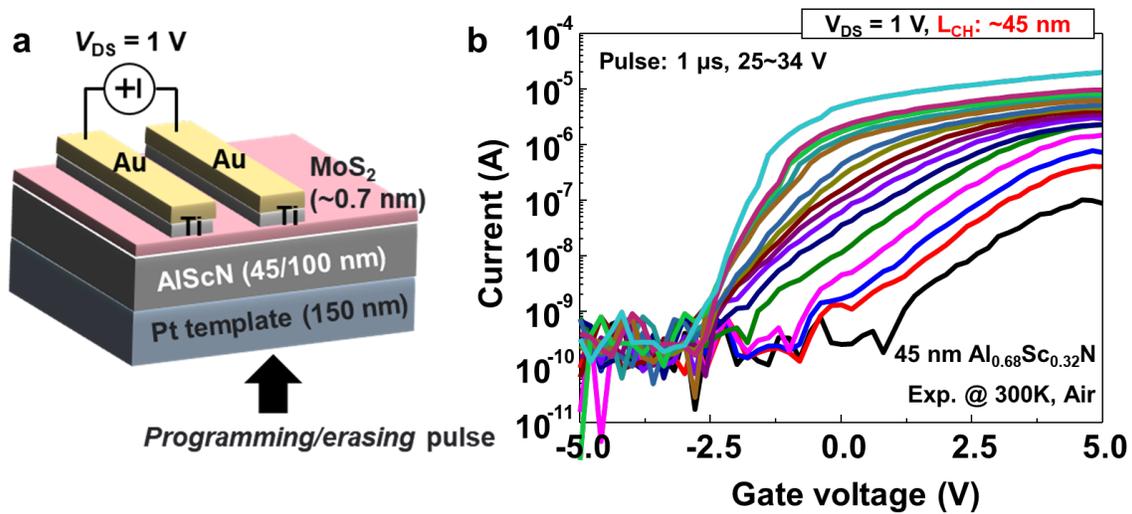

**Fig. S11. 4-bit $V_{TH}$ states of the 45 nm AlScN/MoS$_2$ FE-FETs with $L_{CH}$ of around 80 nm** (a) The programming/erasing pulse schematic. To get the intermediate threshold voltage states, "an amplitude of 7.5 – 12 V and 40 ms" or "an amplitude of 25 – 34 V and 1 µs" was used. (b) The 4-bit operation obtained.

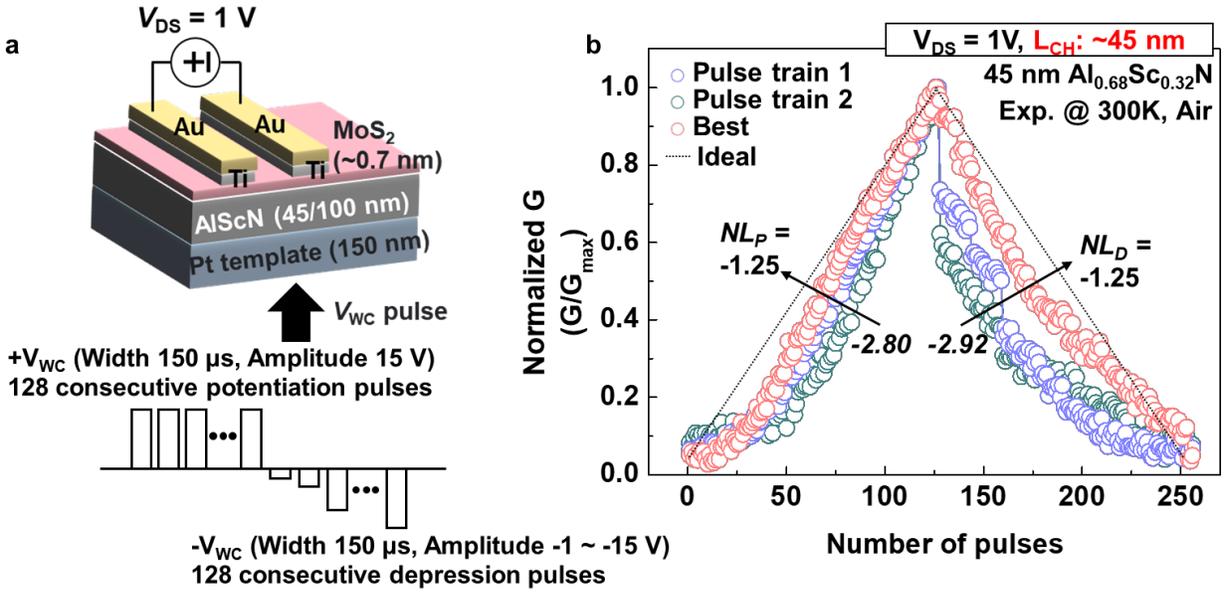

**Fig. S12. Input pulse scheme for high linearity and symmetricity in the FE-FET based synaptic device and corresponding normalized conductance (G).** (a) Schematic diagram of synaptic device based on MoS$_2$/AlScN FE-FET, and input pulse train with the positive and negative weight control pulses ($V_{WC}$). For the LTP, pulses with a constant amplitude of 15 V and width of 150 μs were applied. However, for the LTD, pulses with a varying amplitude from -1 to -15 V were applied to get high linearity and symmetricity. (b) Normalized LTP/D curves of three different pulse train. LTP/D characteristics such as nonlinearity (NL), symmetricity and dynamic range ($G_{max}/G_{min}$) are critical for achieving high accuracy of learning tasks in neural networks. To achieve high linearity, we optimized the LTP/D characteristics by changing the pulse amplitudes. Specifically, lower NL and asymmetricity in LTD was obtained by applying different pulse amplitudes from -1 to -15 V for each range of G.

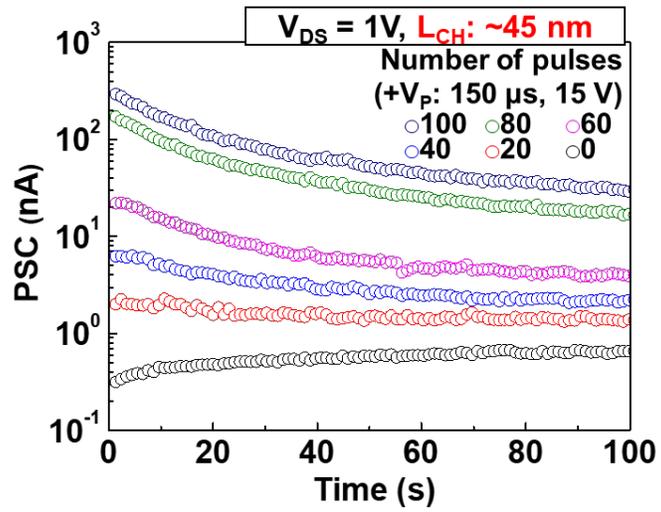

**Fig. S13. The changes of EPSC with the various number of input pulses.** The pulse number ranges from 0 to 100. Positive pulse with the 150 μs width and 15 V amplitude was applied to the gate of FE-FET synaptic device.